\documentclass[10pt,conference]{IEEEtran}
\IEEEoverridecommandlockouts

\usepackage{graphicx}
\usepackage{booktabs}
\usepackage{amssymb}
\usepackage{xcolor}
\usepackage[utf8]{inputenc}
\usepackage[T1]{fontenc}
\PassOptionsToPackage{hyphens}{url}
\usepackage{url}
\usepackage{multirow}
\usepackage{tabularx}
\usepackage{ragged2e}
\usepackage{microtype}
\usepackage{balance}
\usepackage{eso-pic}
\newcommand{\Description}[2][]{}

\begin{document}

\title{CacheFlex: Direct Software-Managed Access to Higher-Level Cache for Scalable Vector Support}

\author{%
\IEEEauthorblockN{Jingqun Zhang\IEEEauthorrefmark{1},
Maohua Nie\IEEEauthorrefmark{1},
Jiayi Wang\IEEEauthorrefmark{1},
Weihang Li\IEEEauthorrefmark{2},
Rishi Sappidi\IEEEauthorrefmark{1},
Shwet Chitnis\IEEEauthorrefmark{1},
Ang Li\IEEEauthorrefmark{1}}
\IEEEauthorblockA{\IEEEauthorrefmark{1}University of Washington, Seattle, WA, USA\quad
\IEEEauthorrefmark{2}Rutgers University, New Brunswick, NJ, USA\\
\{jq86, mhnie, jwang710, rsappidi, s2chitni, angliz\}@uw.edu,\quad
weihang.li@rutgers.edu}
}

\AddToShipoutPictureFG*{%
  \AtPageLowerLeft{%
    \put(50,34){%
      \parbox[b]{512pt}{\fontsize{8}{9.2}\selectfont\raggedright
        \copyright\ 2026 IEEE. Personal use of this material is permitted. Permission from IEEE must be obtained for all other uses, in any current or future media, including reprinting/republishing this material for advertising or promotional purposes, creating new collective works, for resale or redistribution to servers or lists, or reuse of any copyrighted component of this work in other works.}%
    }%
  }%
}
\maketitle

\begin{abstract}
Modern CPUs demand urgent upgrades to keep up with the AI revolution.
While peak compute throughput can be multiplied by widening vector/SIMD lanes, significant improvements of the cache system remain challenging.
We pinpoint L1 data cache capacity as the single design parameter that yields the highest performance and efficiency gains.
However, L1 capacity is constrained by its tight coupling within the core pipeline and the single-digit-cycle latency requirement, so we turn to the private L2 cache that provides an order of magnitude more capacity but is conventionally inaccessible from software.
This work proposes CacheFlex, an ISA and microarchitecture co-design that enables software-managed, direct access to private L2 SRAM by allowing runtime conversion of parts of L2 into a scratchpad memory (SPM).
Through a lightweight ISA extension, software explicitly controls data placement and lifetime within the L2.
CacheFlex integrates into an out-of-order pipeline with cross-domain memory consistency and OS context-switch support, with merely 2.02\% area overhead relative to the L2 in a post-layout 12\,nm implementation.
We prototype CacheFlex based on the Armv8-A SVE ISA, model it in gem5, and evaluate it using optimized general matrix multiplication (GEMM) kernels with diverse model-derived shapes, FlashAttention-based fused attention, and full-layer workloads from LLaMA-3.2-1B and BERT-Base in the prefill phase.
As the vector length (VL) scales from VL\,=\,4 (512-bit) to VL\,=\,16 (2048-bit), CacheFlex's kernel-level geometric-mean speedup and memory-subsystem energy reduction grow from 1.12$\times$ and 26\% to 1.44$\times$ and 40\%, respectively, bridging the widening gap between per-instruction data demand and memory performance.
On full transformer layers combining CacheFlex with FlashAttention, the design delivers up to 1.78$\times$ speedup and 45\% memory-subsystem energy savings (BERT-Base, sequence length $T{=}4096$, VL\,=\,16; LLaMA-3.2-1B: 1.64$\times$ and 38\%).
CacheFlex is complementary to hardware prefetching, and its benefit grows as vector width and peak compute throughput increase for compute-intensive AI kernels.
\end{abstract}


\section{Introduction}
\label{sec:introduction}

Specialized AI accelerators such as GPUs and TPUs dominate the training and serving of large AI models, yet general-purpose CPUs remain indispensable in the AI stack for two reasons.
First, besides orchestrating accelerators and system I/O, CPUs take on many compute- and memory-intensive tasks such as pre-processing multi-modal data, post-processing generated tokens,
and performing non-AI tasks in agentic AI workflows.
Second, CPUs are increasingly compelling for direct AI inference under tight on-device latency, power, and thermal constraints~\cite{edge_ai_market,cpu_gpu_edge} and in on-premises enterprise deployments, where aggregating suitable AI workloads onto CPUs can lower total cost of ownership~\cite{intel_cpu_inference}.
Both roles demand higher
compute throughput from CPU cores.

Modern vector and packed single-instruction, multiple-data (SIMD) architectures leave ample headroom for scaling compute throughput by widening SIMD lanes and replicating SIMD functional units, yet the memory system cannot keep up.
Specifically, CPUs' hardware-managed cache hierarchy optimized for minimal random-access latency is a poor match for AI kernels' highly regular, predictable memory access patterns.

CPUs' memory performance is the result of the complex interplay among the instruction-issuing frontend, the retiring backend, the load/store unit (LSU), the cache hierarchy, and more.
Synergistically upgrading multiple components is extremely challenging for high-performance CPUs that have already been meticulously optimized.
\textbf{In this work, we pinpoint \textit{L1 data (L1D) cache capacity} as the single design parameter that yields the highest performance and efficiency gains (\S\ref{sec:motivation})}.
This is because a larger L1D cache improves data reuse, which raises attainable throughput for dense tensor operations and reduces memory traffic.
Cache bandwidth is less impactful because optimized kernels such as those in the Arm Compute Library (ACL)~\cite{ac} use registers as a scratchpad for the innermost loops, and register bandwidth scales easily with vector width.
Cache latency also has limited impact when it can be effectively hidden by the reorder buffer (ROB) and LSU.

However, directly enlarging the L1D cache is impractical. The L1D cache is typically integrated into the core execution pipeline, so any additional access latency or increased silicon area would require significant changes to the core microarchitecture.
In contrast, a private L2 cache provides an order of magnitude more on-chip storage than the L1D cache, yet it is not directly accessible in software due to hardware-managed cache coherence, virtual memory, and the lack of instruction set architecture (ISA) support.

A question then arises: \textit{can we provide on-demand, software-managed access to a CPU's private L2 cache while maintaining full compatibility with high-performance CPUs' defining features such as speculative, out-of-order (OoO), superscalar execution and operating-system (OS) support?}

\textbf{This work presents an affirmative answer: CacheFlex,
an ISA and microarchitecture co-design that dynamically converts a configurable subset of private L2 cache ways into a software-managed scratchpad memory (SPM).}
\mbox{An ISA extension} with SPM management and access instructions enables explicit software control of data placement and lifetime within the L2 cache.
SPMs are created at runtime, private to a thread, and their L2 usage can be returned to hardware-managed cache mode when SPMs are explicitly deallocated or their owner thread is preempted.
The non-SPM ways remain fully functional as a standard cache.
A dedicated SPM LSU, with its own SPM load/store queue (LSQ), integrates CacheFlex into the OoO pipeline with full speculative/superscalar execution support and OS compatibility.
The dedicated SPM LSU decouples register-facing SPM accesses
(\texttt{spm\_ld}/\texttt{spm\_st}) from cross-domain transfers
(\texttt{spm\_cp}/\texttt{spm\_wb}), allowing data movement to overlap
computation and helping hide L2 latency.
CacheFlex also synergizes well with other cache optimizations such as hardware prefetchers.

\textbf{Scope.}
CacheFlex is a \emph{per-core} memory mechanism, and we evaluate it on the compute-intensive general matrix multiplication (GEMM) and attention kernels that dominate the \emph{prefill} (compute) phase of AI inference.

To summarize, this work makes the following contributions:

\begin{itemize}

\item \textbf{Characterizing the CPU compute--memory gap.}
We show that widening SIMD vectors makes L1 capacity the primary bottleneck of GEMM throughput, driving persistent L1--L2 data movement and energy overhead.

\item \textbf{CacheFlex architecture.}
We propose CacheFlex, an ISA and microarchitecture co-design that enables software-managed, direct access to the private L2 cache. We present a concrete realization on the Armv8-A ISA with Scalable Vector Extension (SVE), though the design principles are ISA-agnostic.

\item \textbf{Evaluating CacheFlex's microarchitecture overheads and performance/efficiency gains across kernels and workloads.}
We synthesize CacheFlex component RTL in a commercial 12\,nm FinFET technology for accurate power and area measurements.
We then model CacheFlex in gem5 and evaluate it (single-core) across a wide range of GEMM dimensions and tiling configurations, fused kernels (FlashAttention-like~\cite{flashattention, dao2023flashattention2}), and full-layer workloads from LLaMA-3.2-1B and BERT-Base in the prefill phase.
At the kernel level, CacheFlex achieves a geometric-mean
1.12$\times$ speedup with 26\% memory-subsystem energy reduction at vector length (VL)\,=\,4
(512-bit), scaling to 1.44$\times$ with 40\% energy reduction
at VL\,=\,16 (2048-bit). On full transformer layers with
FlashAttention, it delivers up to 1.78$\times$ speedup and
45\% memory-subsystem energy savings, at 2.02\% area overhead relative to the L2.

\end{itemize}

\section{Motivation}
\label{sec:motivation}

\subsection{SIMD Throughput Scaling}
\label{sec:compute_memory_gap}

The peak SIMD throughput of a CPU can be raised either by increasing the effective SIMD instructions per cycle (IPC) or by increasing the number of arithmetic operations per instruction (OPs/Inst).
The latter can be achieved more easily by widening the SIMD lanes, thereby attracting more widespread commercial adoption:
Intel AVX-512 provides 512-bit vectors~\cite{avx512}.
Arm SVE defines vector lengths ranging from 128-bit to 2048-bit~\cite{arm_cpu}, and the Fujitsu A64FX processor implements a 512-bit SVE configuration~\cite{a64fx}.
RVV dramatically expands this space with implementation-defined vector lengths up to 65,536 bits~\cite{rvv}. Ara~\cite{ara_tvlsi}, Ara2~\cite{ara2_tc}, and Titan-I~\cite{titan_micro24} demonstrate practical, scalable long-vector processors at different lengths.

However, raising peak throughput without a proportional memory upgrade would result in frequent pipeline stalls due to cache misses, limiting the attainable throughput.
To address this limitation, we must first analyze and pinpoint the primary bottleneck of the CPU's memory hierarchy.

\begin{figure}[tp!]
  \centering
  \includegraphics[width=0.93\linewidth]{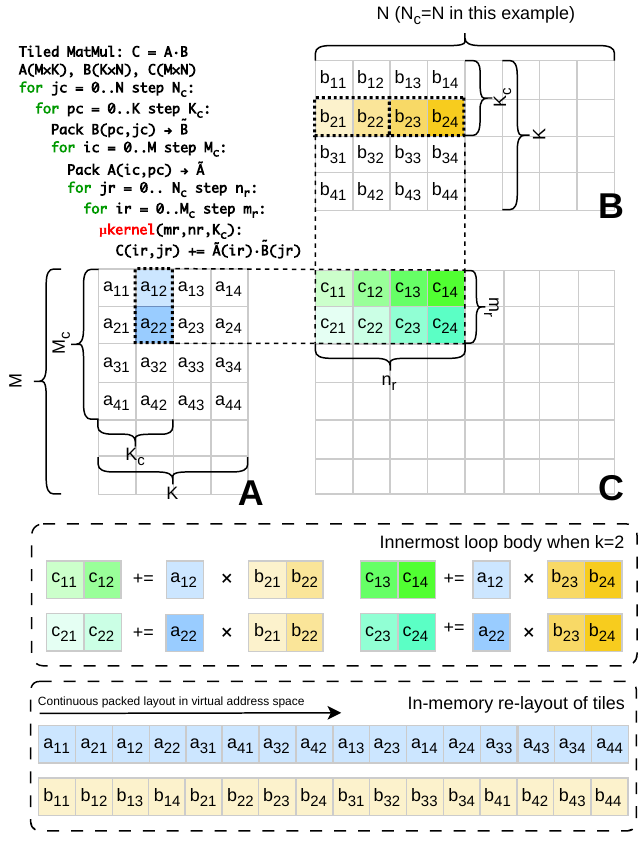}
  \Description{Multi-level GEMM tiling diagram: blocking factors target reuse of the matrix A and B panels at the L1 and L2 cache levels, with the packed contiguous in-memory layout shown below.}
  \caption{Multi-level tiling and packing for an $m_r \times n_r$
  output-stationary GEMM. Outer factors target panel reuse at different cache
  levels ($\tilde{B}$: L1, $\tilde{A}$: L2); the bottom shows the packed
  in-memory layout.}
  \label{fig:motivation_GEMM}
\end{figure}

\subsection{GEMM: Cache Capacity and Tiling}

We use GEMM as a representative kernel due to its dominance in modern AI workloads.
State-of-the-art CPU GEMM kernels, such as those in the assembly-optimized ACL~\cite{ac}, decompose large GEMMs into a hierarchy of tiles, with each tile sized to fit a target cache level for maximal data reuse~\cite{goto2008anatomy,vanzee2015blis}.
As illustrated in Figure~\ref{fig:motivation_GEMM}, tiles are pre-packed or packed ad hoc into contiguous, SIMD-friendly layouts called \textit{panels}; their blocking factors target reuse at different cache levels, with smaller working panels sized for L1 reuse and larger panels for reuse in higher-level caches.
At the innermost level, a small output-stationary tile of $C$ ($m_r \times n_r$, where $n_r$ is a multiple of the vector length) is held in vector registers and accumulated through vector fused multiply-accumulate (VFMA) instructions.

Optimal tiling strategies are sensitive to cache organizations, particularly cache capacities.
Larger capacity allows for bigger panels, which improves data reuse, reduces panel swaps, and therefore reduces total memory traffic and memory-bound stalls.
Widening SIMD lanes has diminishing gains when the optimal tiling strategy hits the capacity and bandwidth limits across cache levels.

\begin{figure}[tp!]
  \centering
  \includegraphics[width=\linewidth]{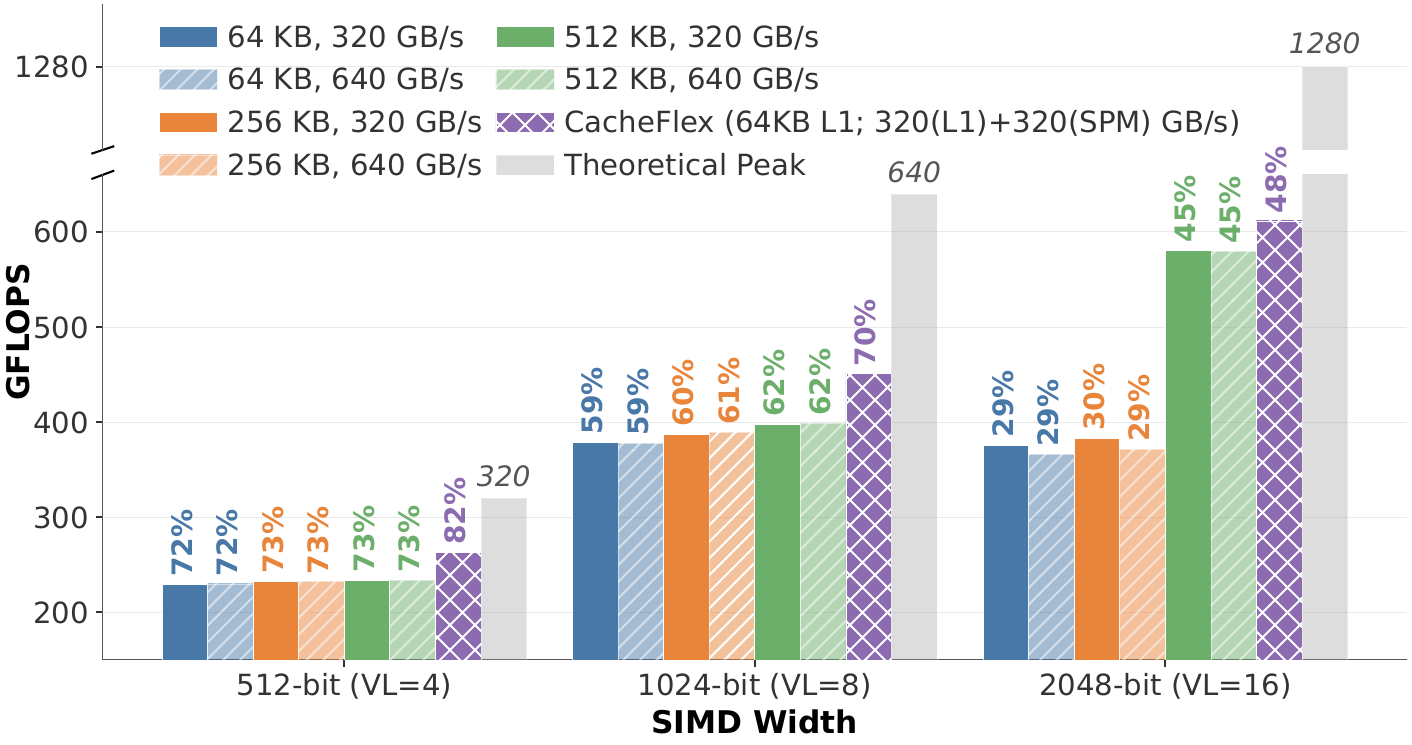}
  \Description{Bar chart of attained GEMM throughput versus SIMD width for L1 capacities from 64 KB to 512 KB at two L1 bandwidths; CacheFlex with a 64 KB L1 matches or exceeds the 512 KB L1 configuration at every SIMD width.}
  \caption{GEMM throughput vs.\ SIMD width and L1D configuration for
  $M{=}256$, $K{=}N{=}2048$. The L2 is fixed at 512\,KB; each configuration
  uses its independently optimized tiling, and annotations show percentage of
  peak throughput.}
  \label{fig:motivation_capacity}
\end{figure}

Figure~\ref{fig:motivation_capacity} confirms this trend quantitatively.
We measure attained compute throughputs of a $256\times2048\times2048$, FP16 GEMM kernel across different L1 data cache capacities and bandwidths.
The innermost compute loop uses an assembly-optimized, $8\times3\mathrm{VL}$ micro-kernel derived from ACL's SVE kernel~\cite{ac}, and each architecture configuration uses its unique optimal tiling strategy identified through a parameter sweep in gem5 simulations.
With a 64\,KB L1, increasing SIMD width from 512-bit (VL\,=\,4) to 2048-bit (VL\,=\,16), a $4\times$ increase in peak throughput, yields only a $1.6\times$ gain in attained throughput: the utilization drops from 72\% of peak throughput at 512-bit to just 29\% at 2048-bit.
Increasing L1 capacity to 512\,KB recovers utilization at 2048-bit to 45\%; beyond that L1 size, the 512\,KB L2 cache capacity becomes the bottleneck.
Critically, doubling L1 bandwidth from 320 to 640\,GB/s yields no measurable improvement across architecture configurations, confirming that the bottleneck is capacity, not L1 bandwidth.
Figure~\ref{fig:motivation_capacity} also previews the effect of CacheFlex: CacheFlex matches or exceeds the 512\,KB L1 configuration at every SIMD width we evaluate,
with only 64\,KB L1 and the same 512\,KB L2 capacities.
This is primarily capacity relief: CacheFlex's total bandwidth (partly sourced from L2) equals that of the highest-bandwidth L1 configuration, and doubling L1 bandwidth alone yields no gain.

\subsection{Opportunity: Direct Access to L2}
\label{sec:cacheflex_proposal}

Most load/store traffic passes through the L1 data cache, which is hard to scale up due to its tight integration with the core pipeline.
Multiplying its capacity at the cost of increased access latency, silicon area, and power consumption would result in a significant negative impact on the CPU's general-purpose performance on OS and non-AI tasks.
Despite continued process scaling, representative Arm~\cite{arm_a710},
AMD~\cite{amd-zen5}, Intel~\cite{intel_golden_cove}, and
RISC-V~\cite{sifive_x390} designs remain confined to only 32--64\,KB of L1D capacity; differing load/store organizations do not eliminate this fundamental capacity wall.

\textbf{However, modern CPUs do have abundant on-chip storage.}
The private L2 cache offers 256\,KB to 2\,MB of on-chip SRAM~\cite{arm_a710,amd-zen5,intel_golden_cove}, several times the capacity of L1.
Even if ROBs and LSQs are greatly deepened to hide the L1-miss/L2-hit latency,
conventional memory accesses still incur address translation and tag lookup;
an L1 miss additionally requires an L2-to-L1 line fill and may evict a resident
line, resulting in significant energy and L2 bandwidth costs.

\textbf{CacheFlex exploits this opportunity and grants the core direct access
to the L2 SRAM while being compatible with all the CPU's
defining features such as out-of-order execution, software-transparent coherence for conventional-memory accesses,
and OS support.}

By exposing the larger L2 SRAM as a software-managed SPM, CacheFlex enables higher-performance tiling strategies.
After a coherent bridge copy, repeated pure-SPM accesses bypass associative tag lookup, address translation, and coherence operations.
This puts the under-utilized L2 bandwidth to better use and reduces power consumption, lowering the risk of thermal throttling during intense computation phases.

\section{CacheFlex Architecture}
\label{sec:arch}
\begin{table*}[t]
    \centering
    \caption{CacheFlex ISA extension. \{\}: SPM address; []: regular address; \#: register index.}
    \label{tab:isa_table}
    \small
    \begin{tabular}{cccl}
        \toprule
        Category & Instruction Name & Parameters & Description \\
        \midrule
        
        \multirow{3}{*}{SPM Management}
        & \texttt{spm\_config} & addr, data & Set CSRs (detailed in Section~\ref{sec:os_compat}) of L2 Cache/SPM \\
        & \texttt{spm\_allocate} & L2-way bitmap & Set selected L2 ways to SPM mode \\
        & \texttt{spm\_deallocate} & L2-way bitmap & Return selected SPM-mode ways to cache mode \\
        
        \midrule
        
        \multirow{4}{*}{SPM Access}
        & \texttt{spm\_ld} & \{spm\_addr\}, \#register & Load data from the SPM to the register \\
        & \texttt{spm\_st} & \#register, \{spm\_addr\} & Store data from the register to the SPM \\
        & \texttt{spm\_cp} & [addr], \{spm\_addr\} & Directly copy data in regular memory space to the SPM \\
        & \texttt{spm\_wb} & \{spm\_addr\}, [addr] & Write data from the SPM back to regular memory space \\
        
        \bottomrule
    \end{tabular}
\end{table*}

CacheFlex allows converting a configurable subset of L2 ways
into a software-managed SPM at runtime.
The SPM offers a thread-private address space that is separate from
the regular memory space (the conventional coherent memory hierarchy).
When converted to SPM mode, each cache line becomes an SPM
\textit{slot} with an SPM address determined by its way and set index.
A slot is one 64-B cache line. A vector access wider than one line spans
multiple slots and is handled as one architectural SPM access; a 2048-bit
access therefore spans four slots.
Once a slot is in SPM mode, pure SPM accesses use its SPM address,
bypassing address translation, associative regular-cache tag matching,
and coherence probes.

\subsection{ISA Extension}
\label{sec:isa}

CacheFlex extends the ISA with a small set of instructions
(Table~\ref{tab:isa_table}) and architectural state
that grant software explicit control over the L2 cache/SPM system.
The instructions fall into two categories.

\textbf{SPM management.}
{\tolerance=1000\emergencystretch=1em
\texttt{spm\_config}, \texttt{spm\_allocate}, and
\texttt{spm\_deallocate} are OS-privileged instructions that
manage the operating mode per L2 way and the privileged control
and status registers (CSRs).
User programs must use system calls to allocate and deallocate SPMs (\S\ref{sec:os_compat}).
\par}

\begin{figure}[t]
  \centering
  \includegraphics[width=0.99\linewidth]{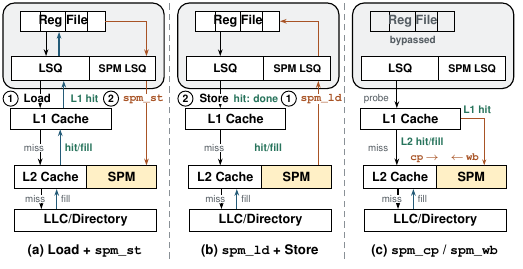}
  \Description{Block diagram of three datapaths between the regular memory space and the SPM: a regular load followed by spm\_st, an spm\_ld followed by a regular store, and an in-cache transfer via spm\_cp or spm\_wb whose transferred data bypasses the register-file data path.}
  \caption{Datapaths between regular memory and SPM. Circled numbers in (a) and
  (b) denote program-visible instruction order; unnumbered arrows show internal
  cache actions. (a) Load then \texttt{spm\_st} and (b) \texttt{spm\_ld} then
  Store route transferred data through the register file. (c) Alternative
  single-instruction bridges \texttt{spm\_cp} (regular to SPM) and
  \texttt{spm\_wb} (SPM to regular) bypass the register-file data path. The
  shown regular-side access paths apply to \texttt{spm\_cp}; \texttt{spm\_wb}
  reuses the normal SQ/coherence write path.}
  \label{fig:shuffle}
\end{figure}

\textbf{SPM access.}
\texttt{spm\_ld} and \texttt{spm\_st} are userland SPM \emph{access}
instructions that transfer data between vector registers and SPM
slots.
\texttt{spm\_cp} and \texttt{spm\_wb} are userland \emph{bridge}
instructions for moving data between the regular memory space and
the SPM memory space, using cache-hierarchy datapaths without incurring round-trip load-to-register
and store-from-register overheads. Figure~\ref{fig:shuffle} shows these datapaths.

\begin{figure}[tp!]
  \centering
  \includegraphics[width=0.78\linewidth]{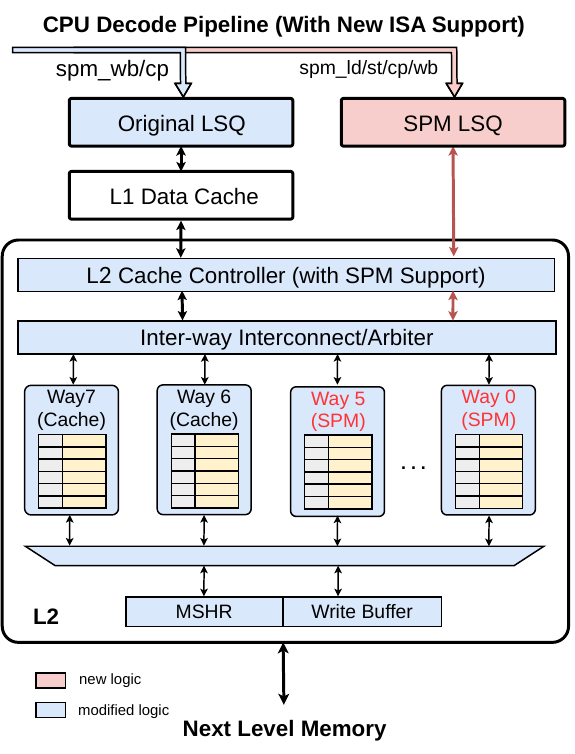}
  \Description{Out-of-order core pipeline connected to an eight-way L2 cache in which ways 0 through 5 operate in SPM mode and the remaining ways stay in cache mode; a dedicated SPM load/store queue sits beside the regular load/store queue.}
  \caption{CacheFlex overview: L2 ways 0--5 operate in SPM mode; ways 6--7
    remain a conventional cache.}
  \label{fig:cacheflex_arch}
\end{figure}

\subsection{Microarchitecture Overview}
\label{sec:arch_overview}

Figure~\ref{fig:cacheflex_arch} overviews the CacheFlex
microarchitecture, with the following added/modified microarchitecture
components in an out-of-order core.

\begin{figure*}[tp!]
  \centering
  \includegraphics[width=0.95\textwidth]{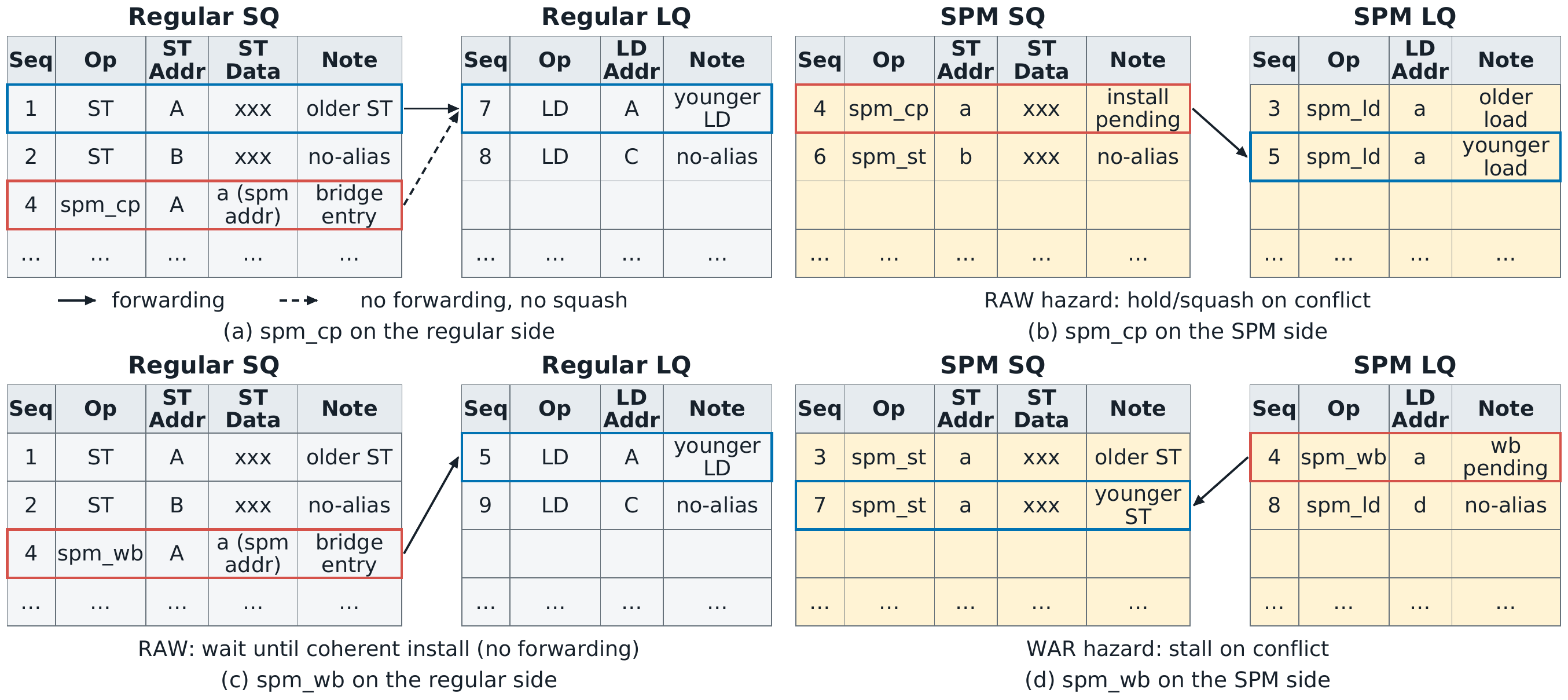}
  \Description{Four queue diagrams showing how spm\_cp and spm\_wb occupy entries in the regular and SPM load/store queues. For spm\_wb, the pending regular write orders younger overlapping accesses without forwarding until coherent installation, while the SPM-side placeholder protects the source slot until its value is captured.}
  \caption{Dual-LSQ coordination for bridge instructions. (a,b) For
  \texttt{spm\_cp}, the regular-side entry never forwards data to younger
  regular loads, and the SPM-side entry holds or replays younger conflicting
  SPM accesses. (c,d) For
  \texttt{spm\_wb}, the pending regular write blocks overlapping younger
  accesses (RAW/WAW), without forwarding, until installation is visible; the
  SPM-side placeholder protects the source until its value is captured (WAR).
  The solid connector in (c) marks a real RAW ordering dependency but carries
  no forwarded data.}
  \label{fig:spmlsq}
\end{figure*}

\textbf{SPM LSQ.}
SPM accesses (\texttt{spm\_ld} and \texttt{spm\_st}) are dispatched exclusively to a dedicated SPM load/store queue (an SPM LQ and SQ), bypassing the regular LSQ entirely.
Because SPM accesses use physical way/slot addressing, the SPM LSQ requires fewer address bits and bypasses associative regular-cache tag matching and address translation.
The SPM LSQ preserves memory ordering by standard address-conflict detection and squash/replay under speculative execution.
Bridge operations (\texttt{spm\_cp} and \texttt{spm\_wb}) touch both memory spaces and are therefore inserted into both the regular LSQ and the SPM LSQ; their cross-space coordination is detailed in Section~\ref{sec:lsq_coord}.

\textbf{L2 cache extensions.}
CacheFlex adds minimal overheads to the L2 cache by reusing existing components as much as possible.
First, the L2 cache controller is extended with lightweight SPM control logic that handles the new SPM instructions and direct SPM-slot accesses.
Bridge instructions (\texttt{spm\_cp} and \texttt{spm\_wb}) are decomposed into coordinated regular- and SPM-side operations that reuse the corresponding datapaths.
In addition, SPM reads use a dedicated, bank-arbitrated L2 access path that avoids contention for the regular L2 request port.
Second, one additional coherence state is added to mark a cache line in SPM mode (\texttt{X} for SPM).
Third, the tag array is overloaded to store the owner thread's ID, letting SPM contents be retained in the L2 across a context switch for opportunistic reuse (\S\ref{sec:os_compat}).

\subsection{Cross-Domain Memory Ordering}
\label{sec:lsq_coord}
Bridge instructions (\texttt{spm\_cp}, \texttt{spm\_wb}) cross the boundary between the regular and SPM address domains, creating potential read-after-write (RAW), write-after-write (WAW), and write-after-read (WAR) hazards between the two sides. CacheFlex handles this by inserting each bridge operation into \emph{both} the regular LSQ and the SPM LSQ, so that each domain independently enforces its own ordering invariants while the bridge instructions maintain consistency at the boundary. Pure SPM operations remain ordered exclusively by the SPM LSQ.
The regular operand of each bridge undergoes normal address translation and
permission checks; a faulting or squashed bridge performs no architectural
destination update.

\subsubsection{\textbf{Coordination for \texttt{spm\_cp} (regular $\to$ SPM)}}
This instruction allocates entries in both the regular store queue (SQ) and the SPM SQ.

  \begin{itemize}
  \item \textit{Regular side (source):}
  \texttt{spm\_cp} pairs a coherent read with an SPM write. The non-destructive, non-allocating source read may be initiated early once its address and older conflicting stores are resolved, while the destination SPM write remains commit-gated.
  It participates in normal SQ ordering with respect to older stores to  the same address; however, because its data is destined for the SPM rather than the regular domain's registers, it does not act as a source for store-to-load forwarding and does not trigger store-violation checks for younger regular loads (Figure~\ref{fig:spmlsq}(a)).

  \item \textit{SPM side (destination):} \texttt{spm\_cp} behaves as a store in the SPM SQ using its \textbf{destination SPM address}. A key challenge is its \emph{variable latency}: fetching data from the regular hierarchy (e.g., due to L2 cache misses) may take hundreds of cycles. The LSQ explicitly manages forward-looking RAW and WAW hazards under this uncertainty while relying on standard instruction retirement for WAR hazards (Figure~\ref{fig:spmlsq}(b)): 
  \emph{(1) RAW Hazard (Looking forward):} Because the \texttt{spm\_cp} data installation can be significantly delayed, the SPM LSQ must actively hold or squash address-matching younger loads to prevent them from reading stale SPM data before the cross-domain transfer completes.
  \emph{(2) WAW Hazard (Looking forward):} Address-matching younger SPM stores remain ordered behind the pending destination entry until installation completes.
  \emph{(3) WAR Hazard (Looking back):} Because the OoO core guarantees in-order commit, \texttt{spm\_cp} will not architecturally update the SPM array until all older instructions, including prior loads, have retired.
\end{itemize}

  \subsubsection{\textbf{Coordination for \texttt{spm\_wb} (SPM $\to$ regular)}}
  \texttt{spm\_wb} mirrors \texttt{spm\_cp}, pairing an SPM read with a coherent write, and reuses the same dual-LSQ coordination. On the SPM side it is a placeholder load at its \textbf{source SPM address}, ordered against same-address SPM stores (Figure~\ref{fig:spmlsq}(d)); on the regular side it is represented as a store at its \textbf{destination regular address} whose write transaction follows the normal SQ/coherence datapath under the same commit-gated discipline (Figure~\ref{fig:spmlsq}(c)). The regular-side entry is not a source for store-to-load forwarding, so overlapping younger accesses wait until coherent installation is visible; unrelated traffic proceeds normally.

\subsubsection{\textbf{Decoupled retirement}}
Bridge instructions retire like \emph{buffered stores}: commit releases the ROB entry without waiting for installation, while destination-side ordering state remains active. For \texttt{spm\_cp}, this state is the destination SPM-SQ entry. For \texttt{spm\_wb}, the regular SQ slot may be released once the coherent path accepts the write transaction, but that path retains a pending destination-address record until the write is installed and visible. The coherent path then acknowledges completion.
Architectural barriers order against this retained state and wait for it when their completion semantics require it.

\noindent
Because SPM regions are thread-private and non-coherent by construction, these rules only need to preserve single-thread memory semantics; no cross-core coherence protocol is required for SPM-resident data. The dual-LSQ coordination enforces address-dependent ordering within each domain and across bridge instructions, reusing the OoO core's existing load/store disambiguation mechanisms.

\subsection{In-Cache Data Transactions}
\label{sec:transactions}

\begin{figure}[t]
  \centering
  \includegraphics[width=\columnwidth]{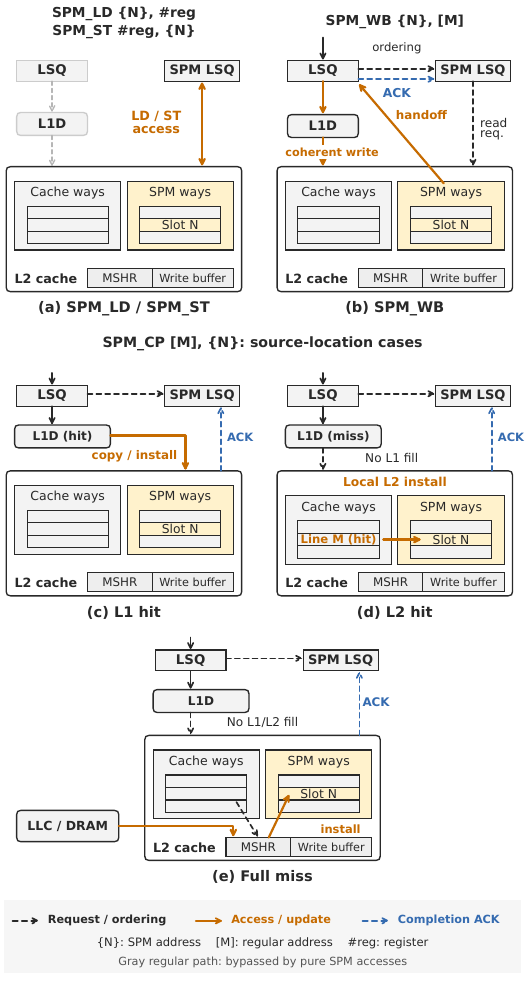}
\Description{Five block diagrams of the L1, L2, and SPM slots showing logical request and ordering coordination, access and update interactions, and completion acknowledgments for spm\_ld and spm\_st, spm\_wb, and spm\_cp under L1-hit, L2-hit, and full-miss cases. A shared legend identifies these component-level interactions, address notation, and the bypassed regular path.}
\caption{In-cache transactions: (a) \texttt{spm\_ld}/\texttt{spm\_st};
(b) \texttt{spm\_wb}; and (c)--(e) \texttt{spm\_cp} source-location cases for
an L1 hit, an L2 hit, and a full miss.}
  \label{fig:isa_detail}
\end{figure}

Data movement between the SPM and regular memory employs a \textbf{lazy-copy policy} to minimize disruption.

\textbf{Demand-driven copy.}
CacheFlex does not pre-populate SPM regions or perform bulk flushes.
Instead, software copies regular-memory data into individual SPM slots on demand using \texttt{spm\_cp}, preventing interconnect traffic bursts and avoiding saturation of the available bandwidth.

\textbf{Non-disruptive state transition.}
While a copy is in progress, the regular hierarchy continues to provide the coherent source value: regular loads and stores can still access it normally, with any necessary ordering enforced by the LSQs (\S\ref{sec:lsq_coord}), not by blocking cache hits.

\texttt{spm\_cp} \emph{copies} a line into the SPM rather than moving it:
it sources data with a non-destructive, non-allocate access to a regular memory address and writes the value into the destination SPM slot.
The source line stays exactly where it was in the regular memory hierarchy, neither invalidated if it hits in any cache nor installed if it misses in any cache.
Because slots convert to SPM mode lazily, the destination slot of the SPM write may still hold a valid cache line; if so, that line is evicted (written back first if dirty) before the conversion completes.

Below we describe how each instruction realizes this
policy. Figure~\ref{fig:isa_detail} illustrates the detailed
transaction behavior; bridge operations obtain or install regular-domain data through the existing coherence protocol, while memory ordering during each transfer is enforced by the dual-LSQ machinery.

\textbf{\texttt{spm\_cp} (regular $\to$ SPM).}
\texttt{spm\_cp} follows the same cache access path as
a regular memory instruction, probing L1 and falling
through to L2 on a miss, to obtain the most up-to-date
version of the target line.
The only extension is that, once the data reaches L2,
it is installed into the target SPM slot via
the L2's existing pipeline
rather than being returned to the register file.
The exact path depends on where the line currently resides:
    \begin{itemize}
    \item \textit{L1 hit} (Figure~\ref{fig:isa_detail}(c)):
    the L1 copy supplies the up-to-date value to the SPM slot; the source line is left resident, neither allocated nor evicted.
    \item \textit{L1 miss, L2 hit} (Figure~\ref{fig:isa_detail}(d)): the resident L2 line is copied directly into the SPM slot through the existing L2 pipeline.
    \item \textit{L1 and L2 miss} (Figure~\ref{fig:isa_detail}(e)): the L2 allocates a miss status holding register (MSHR) and fetches the line from the last-level cache (LLC) or DRAM, and the fill is routed directly into the SPM slot without being installed on the regular memory side.
      \end{itemize}

    \textbf{\texttt{spm\_wb} (SPM $\to$ regular).}
    \texttt{spm\_wb} mirrors \texttt{spm\_cp} by coupling an SPM-side read to a regular-side write whose transaction follows the normal SQ/coherence datapath (Figure~\ref{fig:isa_detail}(b)). This path obtains writable ownership of the destination and coordinates installation of the SPM-sourced value into the coherent line. The existing coherence protocol handles conflicting copies, and completion is acknowledged only after the installation becomes visible. This reuses the core's normal coherent-write machinery rather than adding a separate protocol; inter-core visibility follows the architecture's normal memory-ordering rules.

  \textbf{\texttt{spm\_ld} and \texttt{spm\_st} (Figure~\ref{fig:isa_detail}(a)).}
  Aside from a slot's one-time lazy conversion, \texttt{spm\_ld} and \texttt{spm\_st}
  operate entirely within the SPM domain with no cache interaction: software
  specifies a physical way/slot address, and the SPM control logic accesses the
  corresponding L2 SRAM directly. Because the SPM region is thread-private and
  non-coherent, these accesses bypass virtual address translation, associative regular-cache tag
  lookups, and coherence probes entirely. Memory disambiguation within the SPM
  domain is handled by the SPM LSQ (\S\ref{sec:lsq_coord}).

\subsection{Cache Coherence Compatibility}

CacheFlex is designed to be compatible with standard hardware coherence protocols such as MESI. Under the lazy-copy policy, when certain L2 ways are first converted into SPM mode, the corresponding cache lines are not bulk flushed. When an SPM instruction touches such a slot, the resident cache line is flushed through the standard coherence path: dirty lines are written back and clean lines are invalidated, following the same semantics the protocol uses for a normal eviction.
Deallocation first drains all outstanding SPM accesses, bridge operations, and
retained destination-side ordering state. If preservation is required, live
SPM contents are then spilled; otherwise, they are discarded. The corresponding
ways subsequently return lazily to cache mode without a bulk state-array update:
residual \texttt{X} entries are treated as \texttt{I} once SPM mode is inactive.

\subsection{OS Compatibility}
\label{sec:os_compat}

CacheFlex is designed to remain compatible with conventional OS abstractions, requiring only modest extensions for SPM context management.

\textbf{Architectural state.}
The OS-visible SPM state consists of a small set of CSRs: \texttt{spm\_ctl} (mode and enable), \texttt{spm\_base}/\texttt{spm\_bound} (valid SPM slot range), and \texttt{spm\_backing\_ptr} (a pointer to a fixed-size, page-aligned backing region in regular physical memory for spill/fill across context switches). SPM management operations such as \texttt{spm\_allocate} and \texttt{spm\_deallocate} are privileged; user code accesses CacheFlex through an OS- or runtime-provided interface.

\textbf{Context-switch protocol.}
At a context switch, the core establishes a precise boundary: younger
instructions are squashed, and all older SPM accesses and bridge operations,
including retained destination-side ordering state, drain before the OS saves
or reassigns the SPM state. This ensures that no partially completed SPM access
or cross-domain transfer remains. The OS then saves the thread's SPM CSRs. Depending on the implementation policy, SPM contents may either remain resident in L2 for opportunistic reuse or be written back eagerly to the backing region before the next thread begins execution. The owner-thread ID prevents one thread from accessing another thread's resident SPM data. On an ownership mismatch, the OS resolves the conflict before reassigning the slot; nonresident data can be restored from the thread's backing region when needed.

\section{Experimental Methodology}
\label{sec:methodology}

\textbf{Simulation infrastructure and baseline.}
CacheFlex is evaluated using a detailed cycle-level timing model in
gem5~\cite{gem5} with an Armv8-A out-of-order
core with private L1/L2 caches and a shared L3
(Table~\ref{tab:cpu_config}).
We use a single-core configuration to isolate the first-order effects of CacheFlex on private-hierarchy data placement and movement, without conflating them with shared-resource contention.
The baseline and CacheFlex share the same core, cache, and memory-system parameters;
CacheFlex only adds the ISA support, SPM LSQ structures, and L2 way-level reconfiguration logic, plus a dedicated, bank-arbitrated SPM read port on the private L2 (included in the 2.02\% area), as required by the
proposed design (\S\ref{sec:arch_overview}).
Direct SPM accesses (\texttt{spm\_ld}/\texttt{spm\_st}) use the same
modeled array-access latency as an L2 hit; bridge operations retain the
latency of the regular-memory path they traverse.
CacheFlex allocates L2 ways to the SPM according to application requirements, while the remaining ways continue to operate as a conventional cache.
\begin{table}[t]
    \caption{Configuration parameters of the modeled Armv8-A OoO core, inspired by publicly documented Arm Cortex-A710 characteristics~\cite{arm_a710}.}
    \label{tab:cpu_config}
    \small
    \renewcommand{\arraystretch}{1.2}
    \begin{tabularx}{\linewidth}{|l|>{\raggedright\arraybackslash}X|}
        \hline
        \textbf{CPU} & 5-wide instruction fetch, 5-wide $\mu$Op commit \\
        (@2.5\,GHz)  & 8-wide $\mu$Op issue/dispatch/writeback \newline
                       80 IQ, 32 LQ, 48 SQ, 128 ROB entries \newline
                       128 Int RF, 192 FP RF, 192 Vector RF \newline
                       32 SPM LQ, 48 SPM SQ (same as regular LSQ) \\
        \hline

        \textbf{Functional} & $2\times$Int ALUs \\
        \textbf{Units}      & $2\times$Int vector/FP FUs \newline
                              $2\times$Load + $1\times$Store ports \\
        \hline

        \textbf{L1I / L1D}& Cache Line Size = 64\,B \\
                            & 64\,KB 4-Way LRU \\
                            & Stride Prefetcher with degree 4 \\
        \hline

        \textbf{L2}         & 512\,KB 8-Way LRU, 64\,B line \\
                            & AMPM Prefetcher~\cite{AMPM}, QueueSize 32 \newline
                              Snoop-based cache coherence protocol \\
        \hline

        \textbf{L3}         & 4\,MB 16-Way LRU \\
                            & BOP Prefetcher~\cite{BOP}, QueueSize 32 \newline
                              Snoop-based cache coherence protocol \\
        \hline
        \textbf{DRAM}       & Dual-Channel DDR4-2400 8x8 17-17-17 \\
        \hline
    \end{tabularx}
\end{table}
\begin{table*}[t]
\centering
\caption{Representative GEMM shapes used in the kernel-level evaluation; model names identify shape provenance.
$M$, $K$, and $N$ follow the BLIS convention
($C_{M \times N} = A_{M \times K} \cdot B_{K \times N}$).}
\label{tab:gemm_workloads}
\small
\setlength{\tabcolsep}{4pt}
\renewcommand{\arraystretch}{1.08}
\begin{tabular*}{\textwidth}{@{\extracolsep{\fill}}cllrrrl@{}}
\toprule
ID & Model & Operator & $M$ & $K$ & $N$ & Shape Trait \\
\midrule
W1 & LLaMA-7B~\cite{llama1} & FFN (SwiGLU) & 128 & 4096 & 11008 & Small $M$, large $K{\times}N$ \\
W2 & LLaMA-3-8B~\cite{llama3_8b} & Attn Q/O Proj. & 256 & 4096 & 4096 & Small $M$, large $K$ \\
W3 & LLaMA-3.2-1B~\cite{llama3} & Attn Q/O Proj. & 256 & 2048 & 2048 & Small $M$, medium $K$ \\
W4 & BERT-Base~\cite{bert} & Attn Proj. & 512 & 768 & 768 & Balanced \\
W5 & LLaMA-3.2-1B~\cite{llama3} & Attn Q/O Proj. & 2048 & 2048 & 2048 & Large $M$ \\
W6 & LLaMA-3.2-1B~\cite{llama3} & Attn $QK^{T}$ & 2048 & 64 & 2048 & Large $M$, tiny $K$ \\
W7 & ResNet-50-style~\cite{he2016resnet} & Conv. ($1{\times}1$) & 784 & 256 & 1024 & Medium $M$, small $K$ \\
\bottomrule
\end{tabular*}
\end{table*}

\textbf{Evaluation benchmarks.}
We use FP16 SVE micro-kernels optimized with inline assembly and derived from
ACL~\cite{ac}, with tiling and register blocking
following established high-performance GEMM
techniques~\cite{goto2008anatomy,vanzee2015blis}.
The evaluation has three parts:

\begin{enumerate}
  \item \textbf{Kernel-level GEMM analysis.}
  We evaluate seven GEMM shapes drawn from
  transformer attention, feed-forward layers, and CNN
  pointwise convolutions (Table~\ref{tab:gemm_workloads}),
  covering both compute-bound and memory-sensitive
  regimes.
  For each shape, we sweep tiling strategies
  $(K_c, M_c, N_c)$ (the outer blocking factors defined in
  Figure~\ref{fig:motivation_GEMM}) separately for the baseline and
  CacheFlex across SVE vector lengths
  from 512 to 2048 bits (VL\,=\,4, 8, 16), and report
  the highest-throughput tiling for each design
  point. 

  \item \textbf{Comparison against alternative mechanisms.}
  We compare CacheFlex against representative software and hardware mechanisms for keeping reused data on-chip: hardware prefetching (enabled in the baseline, with a prefetcher-off configuration for contrast), hand-tuned L1 software prefetching, an L2-targeted cache-prefetch hint (\texttt{PLDL2KEEP}~\cite{arm_arm}), and an idealized configuration combining L2 pinning with non-temporal loads. This tests whether CacheFlex's gains are reachable by these simpler means or require its decoupled L2 datapath (\S\ref{sec:capacity_wall}).

\item \textbf{Layer-level inference relevance.}
We evaluate representative layers from LLaMA-3.2-1B~\cite{llama3} and BERT-Base~\cite{bert} during prefilling, including FlashAttention-like fused operator sequences~\cite{flashattention}, to assess whether kernel-level gains propagate to inference-relevant execution contexts. Sequence lengths beyond BERT's standard 512-position limit use BERT-Base--shaped encoder layers to probe capacity scaling. We focus on prefilling because it contains the large dense matrix operations targeted by CacheFlex.
\end{enumerate}

\textbf{FlashAttention implementation.}
CacheFlex enables a more efficient FlashAttention implementation through explicit control over data placement and reuse.
In the CacheFlex variant, software places selected streaming K/V tiles in the SPM, while the remaining operands use the conventional cache hierarchy.
Figure~\ref{fig:layer_eval} reports full-layer results, obtained by combining the simulated time and memory-subsystem energy of all constituent operators according to their model-defined multiplicities.

\textbf{Compilation toolchain.}
All benchmarks are compiled with GCC~15.2 for
AArch64 using \texttt{-O3
-march=\allowbreak armv8.2-a+sve+fp16}.
The computational micro-kernels use optimized inline assembly in
both variants, and each variant uses the best tiling parameters
found by the above sweep.
CacheFlex mnemonics (\texttt{spm\_ld}, \texttt{spm\_st},
\texttt{spm\_cp}, \texttt{spm\_wb}) are translated by a
custom encoding pass inserted between compilation and
assembly; the baseline uses standard SVE memory
instructions (\texttt{ld1h}/\texttt{st1h}) under the
same compiler flags.

\textbf{Energy and area.}
We evaluate energy and area by combining architectural simulation with physical implementation. Per-access energy for the cache hierarchy (L1/L2/L3) and SPM datapath is obtained from RTL implementations synthesized and laid out in a commercial 12\,nm FinFET process.
Gate-level simulation waveforms are analyzed with
Synopsys PrimePower~\cite{ppwr} using switching-activity annotations
to obtain dynamic energy per access and leakage power, and area is taken
from the post-place-and-route layout.
We combine dynamic per-access energy with gem5 access counts and integrate
leakage power over simulated execution time.
DRAM energy is modeled using gem5's integrated power
model based on the configured memory parameters
(Table~\ref{tab:cpu_config}).
We report memory-subsystem energy (cache + SPM + DRAM) as
this captures the dominant effect of CacheFlex, which
primarily reshapes the memory access path.

\textbf{Scope of evaluation.}
Our evaluation is scoped to single-core, mechanism-level analysis, establishing the intrinsic performance and energy implications of CacheFlex where it directly changes execution behavior.
CacheFlex reduces per-access energy on the SPM path by bypassing tag lookup, address translation, and coherence checks.

\section{Results}
\label{sec:results}

\subsection{Kernel-Level GEMM Evaluation}
\label{sec:gemm}

\begin{figure*}[t]
  \centering
  \includegraphics[width=\textwidth]{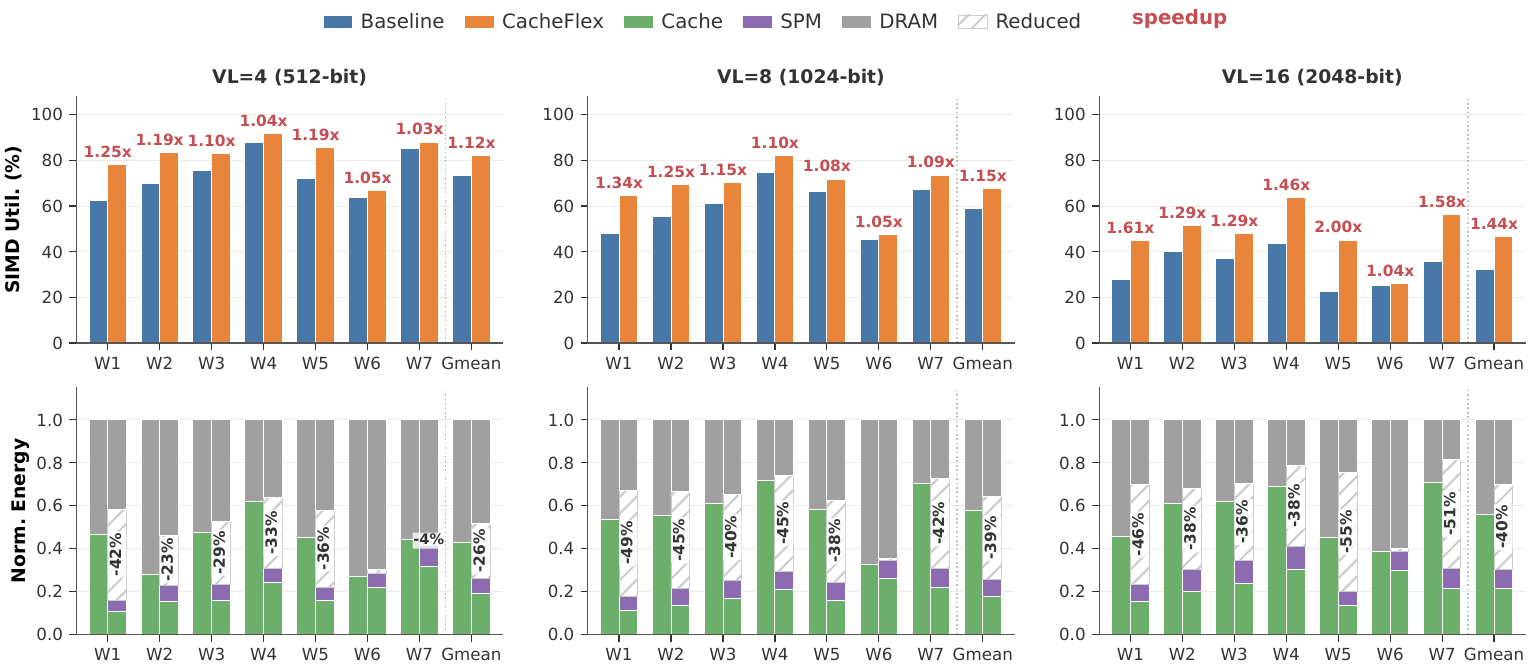}
  \Description{Grouped bar charts of SIMD utilization and normalized memory-subsystem energy for GEMM shapes W1 through W7 at vector lengths 4, 8, and 16, comparing the baseline with CacheFlex.}
  \caption{CacheFlex kernel-level evaluation across GEMM shapes
  (Table~\ref{tab:gemm_workloads}). Top row: SIMD utilization for the baseline
  and CacheFlex across VL\,=\,4, 8, and 16, with speedup annotations. Bottom
  row: normalized memory-system energy for the same configurations, decomposed
  into cache, SPM, and DRAM, with signed annotations for energy changes above
  4\% in magnitude; negative values denote reductions.
  In each pair, the left bar is the baseline and the right bar is CacheFlex;
  hatched regions indicate energy savings, not energy consumption.}
  \label{fig:vl_heatmap}
\end{figure*}

Figure~\ref{fig:vl_heatmap} sweeps SVE vector lengths from VL\,=\,4 to VL\,=\,16 across all seven GEMM shapes and reveals two regimes: a latency-tolerant regime at low VL and a capacity-limited regime at high VL.

\textbf{VL\,=\,4: latency-tolerant regime.}
At VL\,=\,4, the baseline achieves 62--88\% utilization and CacheFlex improves modestly, with geometric-mean speedup and memory-subsystem energy reduction of 1.12$\times$ and 26\%, respectively. In this regime, tiling need not strictly respect cache-capacity bounds because out-of-order scheduling, hardware prefetching, and memory-level parallelism can overlap much of the miss latency. As a result, CacheFlex and the baseline achieve similar throughput, and the main difference is energy: CacheFlex avoids repeated cache-hierarchy overheads by serving data directly from the SPM.
The gain is largest for small-$M$ shapes, where data packing and movement account for a larger fraction of total time; W1--W3 achieve 1.10--1.25$\times$ speedup. By contrast, W6, with $K{=}64$, shows only 1.05$\times$ speedup because its small $K$ keeps the repeatedly reused B-panel compact, leaving little L1 capacity pressure for CacheFlex to relieve.

\textbf{VL\,=\,16: capacity-limited regime.}
As VL increases, operand demand per instruction rises while L1 capacity remains fixed, pushing the baseline into a capacity-limited regime. Figure~\ref{fig:vl_heatmap} shows that baseline utilization drops substantially from VL\,=\,4 to VL\,=\,16 across all shapes, whereas CacheFlex degrades more gradually. Once capacity misses can no longer be hidden, widening VL does not translate into proportional throughput scaling because execution becomes increasingly limited by refill and data-movement overheads rather than compute throughput alone. By providing direct access to software-managed L2 storage, CacheFlex reduces cache pressure and preserves throughput as vector width scales. At VL\,=\,16, the geometric-mean speedup and memory-subsystem energy reduction reach 1.44$\times$ and 40\%, respectively. 
The geometric-mean speedup grows monotonically with VL, though individual shapes may saturate in regimes limited by off-chip memory bandwidth.

\subsection{The Capacity Wall and the Decoupled SPM Datapath}
\label{sec:capacity_wall}
\begin{figure}[tp!]
  \centering
  \includegraphics[width=0.94\columnwidth]{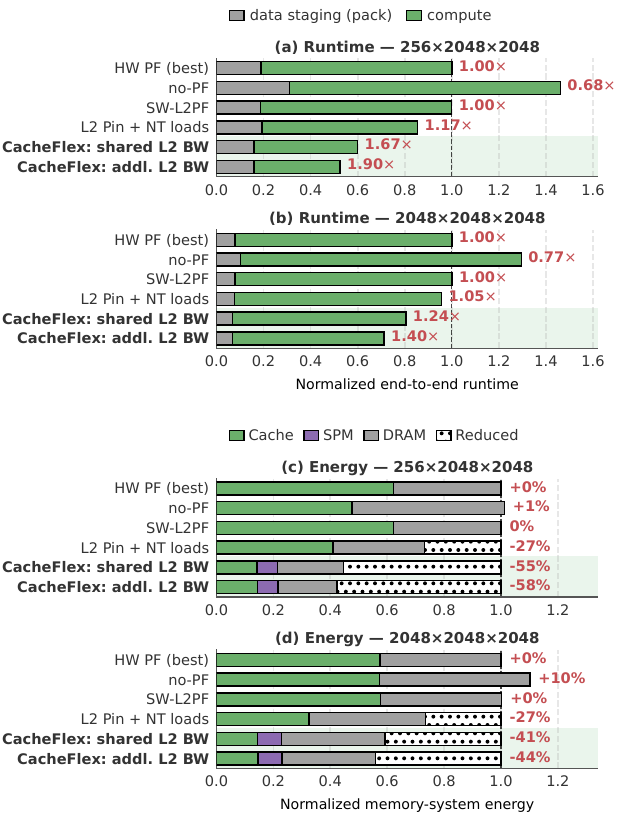}
  \Description{Horizontal bar charts of end-to-end runtime and normalized memory-subsystem energy for CacheFlex and the evaluated software and hardware alternatives on two GEMM shapes, with speedups and energy changes annotated.}
  \caption{CacheFlex vs.\ alternatives for W3 and W5 at VL\,=\,16 on one
  core. (a,b) End-to-end runtime, including staging, normalized to the
  hardware-prefetch baseline; labels give speedup. (c,d) Total memory-system
  energy for the same shapes and baseline; labels give the change.}
  \label{fig:capacity_wall}
\end{figure}

Figure~\ref{fig:capacity_wall} compares CacheFlex against the alternatives shown on two
GEMM kernels, W3 ($256{\times}2048{\times}2048$) and W5
($2048{\times}2048{\times}2048$).
Each alternative uses an independently swept optimal tiling configuration; all runs use VL\,=\,16.
This figure uses a common comparison-oriented kernel organization, whereas Figure~\ref{fig:vl_heatmap} uses independently tuned GEMM implementations; per-shape speedups are therefore comparable within, but not directly across, the two figures.
None matches CacheFlex; we examine each as follows.

\textbf{Software/hardware prefetch.} Hardware prefetch is enabled in
the baseline and already captures GEMM's predictable data flow, giving
$+$46\%/$+$29\% over no prefetch on W3/W5 (Figure~\ref{fig:capacity_wall}(a,b)). With
it enabled, the hand-tuned L1 software-prefetch configurations we evaluated
provide negligible additional benefit over the hardware-prefetch baseline and
are therefore not plotted separately in Figure~\ref{fig:capacity_wall}. Reuse through the L1
is fundamentally bounded by its capacity, so performance is capped by the
resulting L1--L2 swap traffic, which prefetch cannot reduce.

\textbf{L2-targeted prefetch hint (\texttt{PLDL2KEEP}).}
\texttt{PLDL2KEEP} realizes only a subset of \texttt{spm\_cp}: it prefetches lines
into the L2 with a temporal-reuse hint but cannot pin them, so one either risks
conflict evictions from interference among the A/B/C streams or stays
conservative and uses less L2 capacity, giving a suboptimal tiling. Because the L2 is
typically physically indexed and physically tagged, conflicts occur on indices that user software
cannot control. \texttt{PLDL2KEEP} also does not bypass the L1, so it remains
limited by the same L1--L2 traffic. SW-L2PF in
Figure~\ref{fig:capacity_wall} denotes this configuration.

\textbf{Hardware prefetch, L2 pinning, and non-temporal loads.} We
additionally model an idealized variant that augments the hardware-prefetch
baseline with true L2 pinning, setting aside how pinned lines would be unpinned
or interact with the replacement policy. The reused lines are accessed with
non-temporal (NT) loads to prevent L1 installations and, once fetched, are
retained in the L2 by the modeled pinning.
This idealized upper-bound configuration comes closest to CacheFlex. Unlike
\texttt{spm\_ld}, however, NT loads still traverse the L1 pipeline, consuming LSQ
entries and L1 MSHRs and paying full address translation, tag lookup, and
coherence overheads. By taking these accesses off the L1 pipeline entirely, CacheFlex outperforms
even this configuration by \textbf{1.63$\times$}/\textbf{1.34$\times$} on
W3/W5.

\textbf{Ablation: shared vs.\ dedicated L2 read bandwidth.}
The above gains have
two sources: bypassing the L1 pipeline and directly accessing the
larger-capacity L2, and the dedicated L2 read bandwidth. To isolate the
bandwidth contribution, we make SPM-read responses share the regular L2
read-data egress while leaving the rest of CacheFlex unchanged. Even with
shared read bandwidth, CacheFlex
still beats this upper-bound configuration by \textbf{1.43$\times$}/\textbf{1.19$\times$},
confirming that the bulk of the gain comes from bypassing the L1 pipeline and
directly accessing the larger L2; the dedicated read path then contributes a
further \textbf{1.14$\times$}/\textbf{1.13$\times$},
so the two effects stack.

\begin{figure*}[t]
  \centering
  \includegraphics[width=0.94\textwidth]{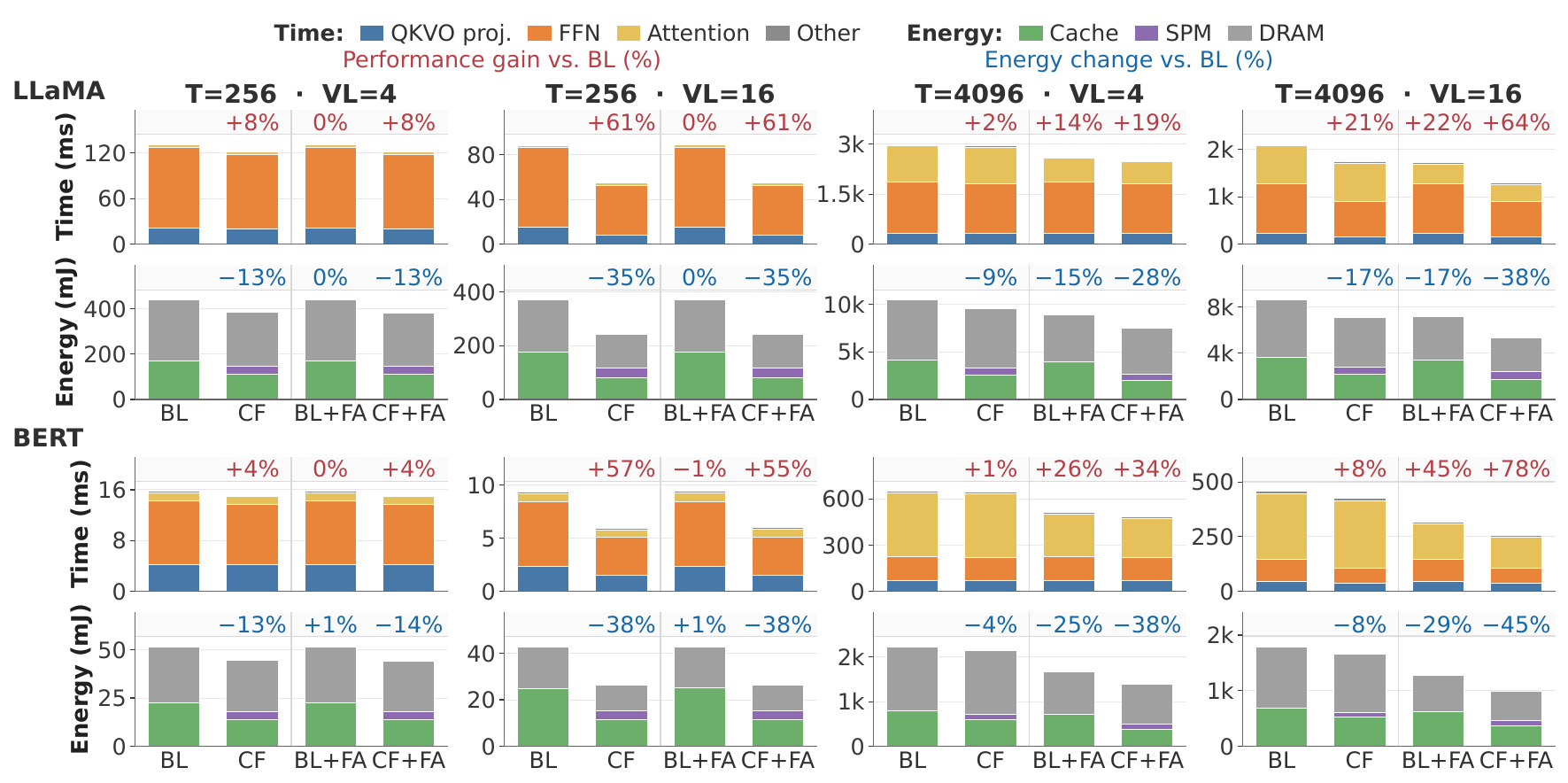}
  \Description{Stacked bar charts of layer execution time and memory-subsystem energy for LLaMA-3.2-1B and BERT-Base prefill layers at sequence lengths 256 and 4096 and vector lengths 4 and 16, under baseline, CacheFlex, FlashAttention, and combined configurations.}
  \vspace{-7pt}
  \caption{Full-layer time and memory-subsystem energy for LLaMA-3.2-1B
  and BERT-Base in the prefill phase, composed as in
  Section~\ref{sec:methodology}. Columns vary $(T,\mathrm{VL})$.
  Red/blue labels show performance gains/energy changes
  relative to BL, rounded to whole percentages; negative blue values indicate
  savings. Other includes normalization, residual operations, and LLaMA
  rotary position embeddings.\newline
  \textbf{BL} = baseline; \textbf{CF} = CacheFlex; \textbf{FA} = FlashAttention;
  \textbf{QKVO}: query/key/value/output projections; \textbf{FFN}: feed-forward
  network; \textbf{SPM}: scratchpad memory; \mbox{\textbf{T}: sequence length};
  \mbox{\textbf{VL}: vector length}.}
  \label{fig:layer_eval}
  \vspace{-9pt}
\end{figure*}

\subsection{Full-Layer Evaluation}
\label{sec:layer}
We evaluate LLaMA-3.2-1B and BERT-Base prefilling layers under four configurations: baseline \mbox{(BL)}, CacheFlex GEMM with unfused attention \mbox{(CF)}, baseline with FlashAttention \mbox{(BL+FA)}, and CacheFlex with FlashAttention \mbox{(CF+FA)}, sweeping $T \in \{256, 4096\}$ and VL $\in \{4, 16\}$.
Figure~\ref{fig:layer_eval} presents $T{=}256$ and $T{=}4096$ as representative endpoints, while Table~\ref{tab:flash_scaling} includes intermediate sequence lengths of $T{=}1024$ and $2048$ to show the behavior as the active per-KV-head $K{+}V$ working set approaches and exceeds L2 capacity.

\textbf{GEMM-dominated regime at $T{=}256$.}
At short sequence lengths, feed-forward network (FFN) and projection GEMMs dominate most of the layer time, so the layer-level result largely reflects kernel-level GEMM behavior.
At VL\,=\,4, CacheFlex GEMM achieves a 1.04--1.08$\times$ speedup over the baseline, since baseline utilization is already high (\S\ref{sec:gemm}) and leaves limited headroom.
At VL\,=\,16, the capacity-miss amplification (\S\ref{sec:gemm}) propagates to the full layer, and CF reaches 1.61$\times$ over BL on LLaMA and 1.57$\times$ on BERT.
At this sequence length, FlashAttention provides little additional layer-level speedup because attention accounts for only a small fraction of total layer time.
In deployment, weights are prepacked at load time, so both BL and CF start from identically formatted tiles.
CacheFlex's throughput advantage therefore reduces to per-access cache-hierarchy overhead, which is modest at VL\,=\,4 but grows with VL (\S\ref{sec:gemm}).
The energy advantage persists regardless of prepacking.

\textbf{Long-sequence regime at $T{=}4096$.}
As sequence length grows, attention accounts for a larger fraction of layer execution time.
FlashAttention alone yields 1.14$\times$ on LLaMA and 1.26$\times$ on BERT at VL\,=\,4.
CacheFlex alone provides only 1.01--1.02$\times$, since the GEMM-side benefit is small at VL\,=\,4 while attention takes a larger share of layer time.
Combining the two reaches 1.19$\times$/1.34$\times$ at VL\,=\,4 and 1.64$\times$/1.78$\times$ at VL\,=\,16 over the baseline.
On LLaMA, the incremental layer-level gain of CacheFlex \emph{on top of} FlashAttention is about 4\% at VL\,=\,4, growing to 34\% at VL\,=\,16, confirming that CacheFlex provides substantial value beyond FlashAttention alone and that this value grows with vector length; BERT shows the same trend.

\textbf{Why CacheFlex amplifies FlashAttention.}
FlashAttention reduces memory traffic by avoiding materialization of the $O(T^2)$ score matrix.
CacheFlex complements this algorithmic benefit with software-controlled residency.
In our CPU implementations, the cache-only variant scans $K/V$ tiles for each query block, whereas the CacheFlex variant stages selected $K/V$ tiles in SPM for reuse across a group of query blocks.
The staged tiles are protected from hardware replacement, while the remaining operands use the cache hierarchy.
This organization reduces repeated data movement and interference with cache-resident operands.

\begin{sloppypar}
\textbf{Contrast: unfused attention.} Without FlashAttention, CacheFlex provides negligible speedup on attention itself.
The bottleneck is the $T{\times}T$ score matrix ($T{=}4096$: 32\,MB), which dwarfs both L2 and SPM.
The constituent GEMMs keep each reused operand tile in L1, so there is no streaming-vs-reuse conflict for CacheFlex to resolve.
\end{sloppypar}

\begin{table}[t]
\centering\small
\caption{FlashAttention kernel: CacheFlex gain (CF+FA vs.\ BL+FA) on LLaMA-3.2-1B under the selected per-shape placement policy. The reported $K{+}V$ footprint is per KV head. Positive entries report measured gains; 0 indicates that SPM mode offers no measurable benefit.}
\label{tab:flash_scaling}
\begin{tabular}{rrrll}
\toprule
$T$ & \shortstack{per-head\\$K{+}V$} & \shortstack{per-head\\$K{+}V$/L2} & VL\,=\,4 & VL\,=\,16 \\
\midrule
256  & 64\,KB   & 12.5\%  & $+3.2\%$  & 0 \\
1024 & 256\,KB  & 50\%  & 0 & 0 \\
2048 & 512\,KB  & 100\% & $+7.4\%$  & $+18.3\%$ \\
4096 & 1024\,KB & 200\% & $+9.7\%$ & $+15.2\%$ \\
\bottomrule
\end{tabular}
\end{table}

Table~\ref{tab:flash_scaling} isolates the FlashAttention kernel-level gain across sequence lengths.
While the per-KV-head $K{+}V$ working set fits comfortably in L2 ($T \leq 1024$), the deployed gain is at most 3.2\%; for shapes where SPM mode offers no measurable benefit, the placement policy retains conventional cache mode.
At $T{=}2048$, this working set is 512\,KB, equal to the L2 capacity; CacheFlex improves performance by $7.4\%$ at VL\,=\,4 and $18.3\%$ at VL\,=\,16.
At $T{=}4096$, where it is twice the L2 capacity, the gain remains positive at $9.7\%$ and $15.2\%$, respectively.
The maximum measured kernel gain is 18.3\%, observed once the working set approaches or exceeds L2 capacity.

\textbf{Energy.}
CacheFlex consistently reduces memory-subsystem energy by shifting accesses from the tag/TLB/coherence cache path to direct SPM reads.
At VL\,=\,4, CF+FA reduces total memory-subsystem energy on LLaMA by 13.1\% at $T{=}256$, growing to 28.3\% at $T{=}4096$; at VL\,=\,16, the reduction grows from 34.6\% at $T{=}256$ to 38.4\% at $T{=}4096$.
Even when throughput gains are modest, such as 1.08$\times$ at VL\,=\,4 and $T{=}256$, energy savings remain substantial, demonstrating that CacheFlex's value extends beyond performance to energy efficiency.

\subsection{Area Overhead}

We estimate CacheFlex's area overhead using post-place-and-route results for a commercial 12\,nm FinFET process. CacheFlex adds two structures to the baseline core: an SPM load/store queue with 32 load entries and 48 store entries, occupying 7{,}084~$\mu$m$^2$, and L2 SPM-related control logic, occupying an additional 1{,}293~$\mu$m$^2$. Together, these structures contribute 8{,}377~$\mu$m$^2$ of additional silicon. CacheFlex introduces no new SRAM arrays; it repurposes existing L2 capacity at runtime, so the main storage cost is already accounted for in the baseline design. As a result, the L2 cache area increases only from 414{,}677~$\mu$m$^2$ to 415{,}970~$\mu$m$^2$, corresponding to a 0.31\% increase. When including the SPM LSQ, the total area overhead relative to the L2 baseline is only 2.02\%.

\begingroup

\section{Related Work}
\label{sec:related_work}

\textbf{Dedicated and physically separate scratchpads.}
Software-managed memories have a long history: Arm tightly coupled memory (TCM)~\cite{arm_tcm}, the Cell Broadband Engine's per-SPE local stores~\cite{gschwind2006chip}, MIT Raw's per-tile 96-KB software-managed instruction cache~\cite{taylor2002raw}, the TPU's unified buffer~\cite{tpu}, and Eyeriss's global buffer~\cite{eyeriss}. These designs gain explicit control by paying for dedicated, statically provisioned on-chip storage rather than reclaiming capacity from an existing cache on demand. The SHADE infrastructure~\cite{shade_latte24} likewise targets distributed scratchpad memories.

\textbf{Cache--scratchpad integration.}
The closest designs to CacheFlex share one SRAM between the cache and a scratchpad. Modern NVIDIA GPU per-streaming-multiprocessor (SM) shared memory~\cite{cuda-shared-memory} splits the on-chip SRAM this way, but targets a SIMT execution and GPU context-management model rather than speculative out-of-order CPU execution.
In the evaluated CPU--GPU system, Stash~\cite{stash} replaces the per-CU scratchpad at the L1 level, providing direct addressing and coherent global visibility. CacheFlex instead reclaims private-L2 capacity for a non-coherent, thread-private SPM and connects it to coherent regular memory through explicit bridge operations.
VLS~\cite{VLS} comes closer by using a dynamically sized L1-data-cache partition as an OS-virtualized local store for the currently running thread, but relies on a user-level DMA engine and was evaluated on a simple in-order core.
Prior hybrid cache--scratchpad designs provide complementary precedents. Kalokerinos et al.~\cite{kalokerinos2019configurable} divide local SRAM at run time between an L2 cache and directly addressed scratchpad regions; their FPGA prototype centers on in-order MicroBlaze cores and virtualized user-level RDMA rather than a coherent out-of-order CPU memory system. Cong et al.'s Adaptive Hybrid Cache~\cite{cong2011adaptive} reconfigures part of an L1 hybrid cache as SPM and dynamically remaps SPM blocks across cache sets. SPX64~\cite{singh2021spx64} places a separate non-coherent scratchpad alongside an out-of-order CPU and provides OS management support. CacheFlex differs in combining reclaimed private-L2 capacity, explicit bridge operations between coherent regular memory and the non-coherent SPM, hardware-managed dual-LSQ cross-domain ordering, and wide-vector compute-side reuse.

\setlength{\parskip}{7.5pt}
\textbf{Runtime- and boot-time reconfigurable memories.}
A broader line of work explores configurable on-chip memory organizations: Smart Memories~\cite{smartmemories2000} configures SRAM \emph{mats} as caches, scratchpads, or stream-register files to match application requirements; the evaluated Reconfigurable Caches design~\cite{reconfigcaches2000} repartitions at cache-way granularity under software control but keeps every access on the tag-checked, associatively indexed path; TI's C64x DSPs~\cite{ti_c64x} repartition the unified L2 between mapped SRAM and cache using a core-wide choice among fixed cache sizes; and Knights Landing~\cite{knl2016} exposes its separate on-package MCDRAM in coarse Flat, Cache, or Hybrid modes. At runtime, CacheFlex converts selected private-L2 ways into an untagged, directly addressed scratchpad managed as per-thread state.

\balance

\begin{sloppypar}
\textbf{Cache partitioning and bypass.}
Intel CAT~\cite{intel_cat}, Arm MPAM~\cite{arm_mpam}, Jigsaw~\cite{jigsaw}, and Jenga~\cite{jenga} can partition or constrain allocation within shared-cache capacity at workload or data granularity, improving isolation and utilization. These mechanisms control cache capacity or placement while leaving individual lines replacement-managed; CacheFlex instead provides software-controlled residency for selected reused tiles. These mechanisms remain orthogonal: a system can partition the shared LLC for isolation while CacheFlex deterministically preserves reused operands in its private L2.

Cache compression (BDI~\cite{bdi}, C-Pack~\cite{cpack}) enlarges effective capacity but adds decompression on the access path, while non-temporal hints (x86 \texttt{movnt*}~\cite{intel_sdm}, Arm \texttt{LDNP}/\texttt{STNP}~\cite{arm_arm}) can reduce pollution but still provide no residency guarantee for reused data.
CacheFlex instead guarantees residency: a per-thread slice of the private L2 becomes a directly addressed region.
\end{sloppypar}

\textbf{Prefetching and decoupled architectures.}
Hardware~\cite{stride_prefetch,AMPM,BOP} and software~\cite{mowry1992prefetch} prefetchers fetch into the ordinary cache to hide latency, but their lines stay replacement-eligible with \textbf{no residency guarantee};
cache pinning/locking~\cite{mittal2016cachelocking} prevents selected lines from being replaced but still pays the latency and energy of the ordinary tagged cache-lookup path.
Prodigy~\cite{prodigy} issues non-binding prefetches into the ordinary L1D, leaving prefetched application data without a residency guarantee. UVE~\cite{unlimited} maps configured streams to vector registers and streams operands directly into the register file, but exposes no separately addressable cache-capacity region.

AMU~\cite{wang2024} partitions L2 into a software-managed scratchpad on an out-of-order core, uses software disambiguation for conflicts among asynchronous accesses, and focuses on hiding far-memory access latency. CacheFlex instead targets compute-side operand reuse, provides hardware disambiguation of cross-domain hazards through dual LSQs, and virtualizes SPM state per thread.

Prefetch and streaming engines remain complementary: they can stage source data
in the coherent hierarchy, after which \texttt{spm\_cp} copies it into the SPM;
CacheFlex then guarantees residency for reused data.

\textbf{Algorithm-level tiling and near-data processing.}
FlashAttention~\cite{flashattention} shows that explicit control over data placement can substantially reduce memory traffic, but mainstream CPUs expose no general residency control, so tiles in active reuse can still be silently evicted.

Near-data designs are orthogonal and memory-side: IMPICA~\cite{impica_iccd16} performs pointer chasing in 3D-stacked memory, Gather-Scatter DRAM~\cite{gsdram_micro15} provides in-DRAM address translation and gather/scatter support for non-unit-stride accesses, and Instant-NeRF~\cite{instantnerf_dac23} uses algorithm--accelerator co-designed near-memory processing. These approaches compose with CacheFlex's on-chip capacity management.

\section{Conclusion}
\label{sec:conclusion}

Modern OoO CPUs can tolerate some L1-capacity overflow through latency hiding, but this tolerance weakens as vector length scales and per-instruction data demand outgrows fixed L1 capacity. In this regime, the key limitation is no longer compute throughput alone, but the conventional cache hierarchy's inability to provide direct, software-controlled access with guaranteed residency in the private L2.

CacheFlex addresses this gap by repurposing existing private L2 SRAM as a software-managed scratchpad, while preserving OoO execution support and incurring only 2.02\% area overhead relative to the L2. 
Across both GEMM and FlashAttention-based workloads, CacheFlex improves memory-subsystem energy efficiency and,
in the capacity-limited high-vector-length regime, performance, by removing cache-management overheads and giving software explicit control over residency.

Overall, exposing existing L2 capacity as a software-managed resource is a practical way to sustain CPU efficiency as vector throughput scales. The approach likely generalizes to capacity-sensitive domains like graph analytics and sparse linear algebra, could extend to bandwidth-bound kernels and emerging matrix engines, and invites OS-level SPM scheduling across threads.

\par
\endgroup

\clearpage
\nobalance
\IEEEtriggeratref{40}
\IEEEtriggercmd{\balance}
\bibliographystyle{IEEEtran}
\bibliography{refs}

\clearpage
\nobalance
\appendix[Artifact Appendix]
\label{app:artifact}

\subsection{Availability and Requirements}
\noindent\url{https://github.com/JingqunZhang/cacheflex-ae}

The artifact includes the CacheFlex gem5 model, instruction support,
benchmarks, experiment scripts, and paper measurements. Inputs are generated
deterministically; no external data sets or model weights are needed.
Use an x86-64 Linux host with Python 3.11 or 3.12.

\subsection{Generate the Paper Figures}
From the repository root, run:

{\small
\begin{verbatim}
python3 -m venv .venv
source .venv/bin/activate
python -m pip install -r requirements.txt
bash reproduce.sh
\end{verbatim}
}

This generates PDFs, PNG previews, JSON data, and Table~\ref{tab:flash_scaling}
in \path{results/} from the supplied measurements, without simulation.

\subsection{Rerun the Simulations}
Full reruns also require GCC/G++ 10--14, \texttt{make}, and Python/zlib
development headers. We recommend 20\,GB of free disk space.
Follow \texttt{SETUP.md} to install Arm GNU Toolchain 15.2.rel1 and
QEMU 7.2.0, build gem5, and run the QEMU and gem5 smoke tests.
These check functional outputs and SPM activity; both should print
\texttt{SMOKE TEST PASSED}.
Then run:

{\small
\begin{verbatim}
source setup_env.sh
bash run_all.sh -j 8
\end{verbatim}
}
Allow roughly 1--2 days at \texttt{-j 8} for the selected configurations,
excluding initial setup; runtime varies by host. Each group's
\texttt{run.sh} also runs independently. Optional \texttt{-{}-sweep} searches
alternative tilings and takes substantially longer; it is not needed to
reproduce the paper results. Use \texttt{-j N} (1--16) to adjust parallelism.
Plots and tables go to \path{results/rerun/}; logs and statistics remain in
each group's \path{results/generated/}.

\subsection{Expected Results and Validation}
{\small
\renewcommand{\arraystretch}{1.12}
\setlength{\tabcolsep}{4pt}
\noindent\begin{tabularx}{\linewidth}{@{}>{\raggedright\arraybackslash}p{0.24\linewidth}>{\raggedright\arraybackslash}X@{}}
\toprule
\textbf{Paper / file} & \textbf{Expected result} \\
\midrule
Fig.~\ref{fig:motivation_capacity}\newline\texttt{fig2.pdf} &
At VL\,=\,16, L1 scaling raises utilization from 29\% to 45\%; CacheFlex reaches 48\%. \\
\addlinespace[3pt]
Fig.~\ref{fig:vl_heatmap}\newline\texttt{fig7.pdf} &
At VL\,=\,16, geometric-mean speedup is 1.44$\times$, with 40\% lower memory-subsystem energy. \\
\addlinespace[3pt]
Fig.~\ref{fig:capacity_wall}\newline\texttt{fig8.pdf} &
CacheFlex achieves 1.90$\times$/1.40$\times$ speedup over HW-PF on W3/W5. \\
\addlinespace[3pt]
Fig.~\ref{fig:layer_eval}\newline\texttt{fig9.pdf} &
At $T{=}4096$, VL\,=\,16, CF+FA gives 1.64$\times$/1.78$\times$ layer speedup over BL on LLaMA/BERT. \\
\addlinespace[3pt]
Table~\ref{tab:flash_scaling}\newline\texttt{table4.tex} &
FlashAttention-kernel throughput gain reaches 18.3\% for CF+FA versus BL+FA. \\
\bottomrule
\end{tabularx}\par
}
\medskip

JSON comparisons against the supplied measurements should print
\texttt{RESULT COMPARISON PASSED}; the wrapper then finishes with
\texttt{DONE}.

\end{document}